\documentclass[prl,amsmath,amssymb,twocolumn,showpacs,letterpaper,superscriptaddress]{revtex4}
\usepackage{graphicx}
\begin{document}
\title{Coexistence of superconductivity and ferromagnetism in two dimensions}

\author{D.A. Dikin}
\affiliation{Department of Physics and Astronomy, Northwestern University, Evanston, IL 60208, USA}
\author{M. Mehta}
\affiliation{Department of Physics and Astronomy, Northwestern University, Evanston, IL 60208, USA}
\author{C.W. Bark} 
\affiliation{Department of Materials Science and Engineering, University of Wisconsin-Madison, Madison, WI 53706, USA}
\author{C.M. Folkman} 
\affiliation{Department of Materials Science and Engineering, University of Wisconsin-Madison, Madison, WI 53706, USA}
\author{C.B. Eom} 
\affiliation{Department of Materials Science and Engineering, University of Wisconsin-Madison, Madison, WI 53706, USA}
\author{V. Chandrasekhar}
\email{v-chandrasekhar@northwestern.edu}
\affiliation{Department of Physics and Astronomy, Northwestern University, Evanston, IL 60208, USA}

\date{\today}


\begin{abstract}
Ferromagnetism is usually considered to be incompatible with conventional superconductivity, as it destroys the singlet correlations responsible  for the pairing interaction.  Superconductivity and ferromagnetism are known to coexist in only a few bulk rare-earth materials.  
Here we report evidence for their coexistence in a two-dimensional system: the interface between 
two bulk insulators, LaAlO$_3$ (LAO) and SrTiO$_3$ (STO), a system that has been studied intensively recently.  Magnetoresistance, Hall 
and electric-field dependence measurements suggest that there are two distinct bands of charge carriers that contribute to the interface 
conductivity.  The sensitivity of properties of the interface to an electric field make this a fascinating system for the study of the interplay between superconductivity and magnetism.
\end{abstract}

\maketitle

There has been much interest recently in the conducting interface that forms between the two band insulators, 
SrTiO$_3$ (STO) and LaAlO$_3$ (LAO) \cite{ohtomo,thiel,reyren,brinkman,ariando,siemons,cen,schneider,huijben,caviglia2,shalom,
seri,caviglia}.  This interface shows a rich variety of behavior, including superconductivity \cite{reyren,caviglia2,shalom}, 
magnetism \cite{brinkman,ariando,huijben,seri}, and electric field controlled metal-insulator \cite{thiel,cen} and superconductor-insulator 
transitions \cite{caviglia2,schneider}.  Two important mechanisms have been proposed for the creation of the conducting layer at the 
interface \cite{siemons,huijben,chen,kalabukhov,seo,fix}: charge transfer from the LAO to the Ti$^{+2}$ ions at the interface 
(the so-called ``polar catastrophe'' mechanism);  and conduction due to oxygen vacancies, which can be controlled by growth or post-growth 
annealing conditions.    The Ti bands are also thought to contribute to the magnetism seen in some samples \cite{chen}.  

The electronic 
characteristics are very sensitive to the growth conditions:  generally, it is found that samples grown in an environment with low oxygen partial pressure 
$P_{O_2}$ have more oxygen vacancies and are consequently more conducting; the conductivity is reduced if the samples are grown
in high $P_{O_2}$, or subjected to a post-growth oxygen anneal \cite{huijben,ariando}.  It is not only the conductivity that is sensitive 
to growth conditions:  superconductivity is observed in samples grown in intermediate $P_{O_2}$ \cite{huijben}, and signatures of 
ferromagnetism are observed for samples grown in high $P_{O_2}$ \cite{brinkman,ariando}, although both phenomena have not been observed 
until now in the same sample.  Here we report measurements on LAO/STO interface structures where both phases are seen simultaneously at 
low temperatures.  Magnetoresistance and Hall measurements indicate that there are at least 
two bands of charge carriers in the system.  Earlier theoretical calculations \cite{pentcheva} have pointed to a ferromagnetic ground state of the system due to the multivalent nature of the Ti ions at the interface, but the origin of superconductivity is still not clear.  This system joins only 
a few other bulk materials in which superconductivity and ferromagnetism have been observed simultaneously 
\cite{felner,saxena,aoki,pfleiderer}, with two critical differences:  both the superconductivity and the magnetism are confined to a 
two-dimensional interface, and the electronic properties of this interface can be tuned over a wide range by means of an electric field.  
Consequently, it forms a unique system for the investigation of the interplay between superconductivity and magnetism.

The films in this work had 10 unit cells (uc) of LAO grown by pulsed laser deposition at $P_{O_2}=10^{-3}$ mbar on TiO$_2$ terminated 
(001) STO single crystal substrates \cite{park}.  For electrical measurements, a Hall bar geometry was patterned by photolithography and etched using 
argon ion milling that removed the LAO layers and a few layers of STO.  Measurements confirmed that the bare etched STO was not conducting from room temperature down to millikelvin temperatures.  A gate voltage $V_g$ applied to the back of the substrate was used 
to modulate the conductance of the devices.  While the qualitative behavior of all the samples 
was the same, the devices showed small sample-to-sample variations that were only evident at millikelvin temperatures.  The origin of these variations is not clear at the moment.  
We shall concentrate in this paper on data from a single longitudinal section and its adjacent Hall configurations that showed the sharpest 
superconducting transitions.     

\begin{figure}[h]
\includegraphics[width=8.7cm]{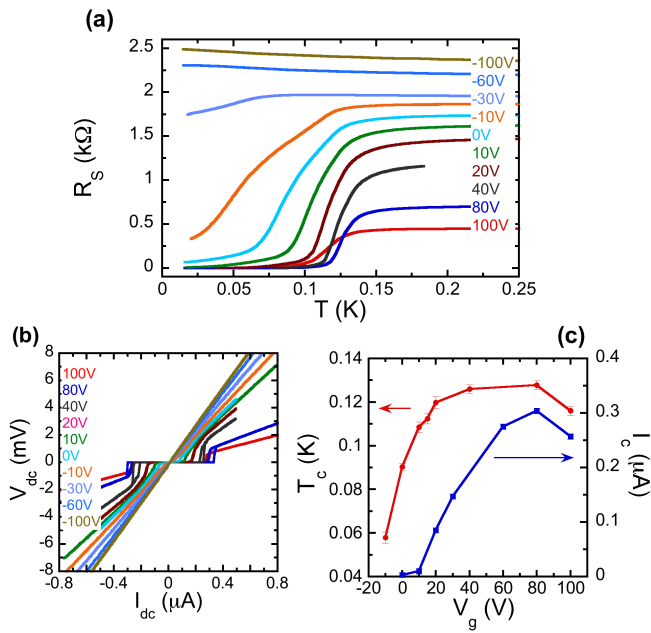}
\vspace{-0.5cm}
\caption{(a) Superconducting transition at different gate voltages $V_g$.   (b)  
Current voltage characteristics of the sample at 15 mK.  (c)  Transition temperature $T_c$ (defined as the midpoint of the resistive 
transition) and critical current (defined as the current at which the sample switches to the resistive state on ramping the current up 
from zero) as a function of $V_g$.  Both measures indicate that the maximal superconducting properties are obtained for $V_g\simeq80$ V.  
Error bars indicate the difference in $T_c$ measured between cooling and warming traces. }
\vspace{-0.5cm}
\label{fig1}
\end{figure}  

The normal state and superconducting characteristics and their dependence on $V_g$ and temperature $T$ are similar to that seen by other groups \cite{reyren,caviglia2,schneider} (Fig. 1(a)).  The current-voltage curves also show a characteristic superconducting signature with a 
critical current $I_c$ that vanishes when $V_g<-20$V (Fig. 1(b)).  However, $I_c$  and the transition temperature $T_c$ do 
not increase monotonically with increasing $V_g$, but show maxima at $V_g \simeq 80$ V (Fig. 1(c)). 

\begin{figure}[h]
\includegraphics[width=8.7cm]{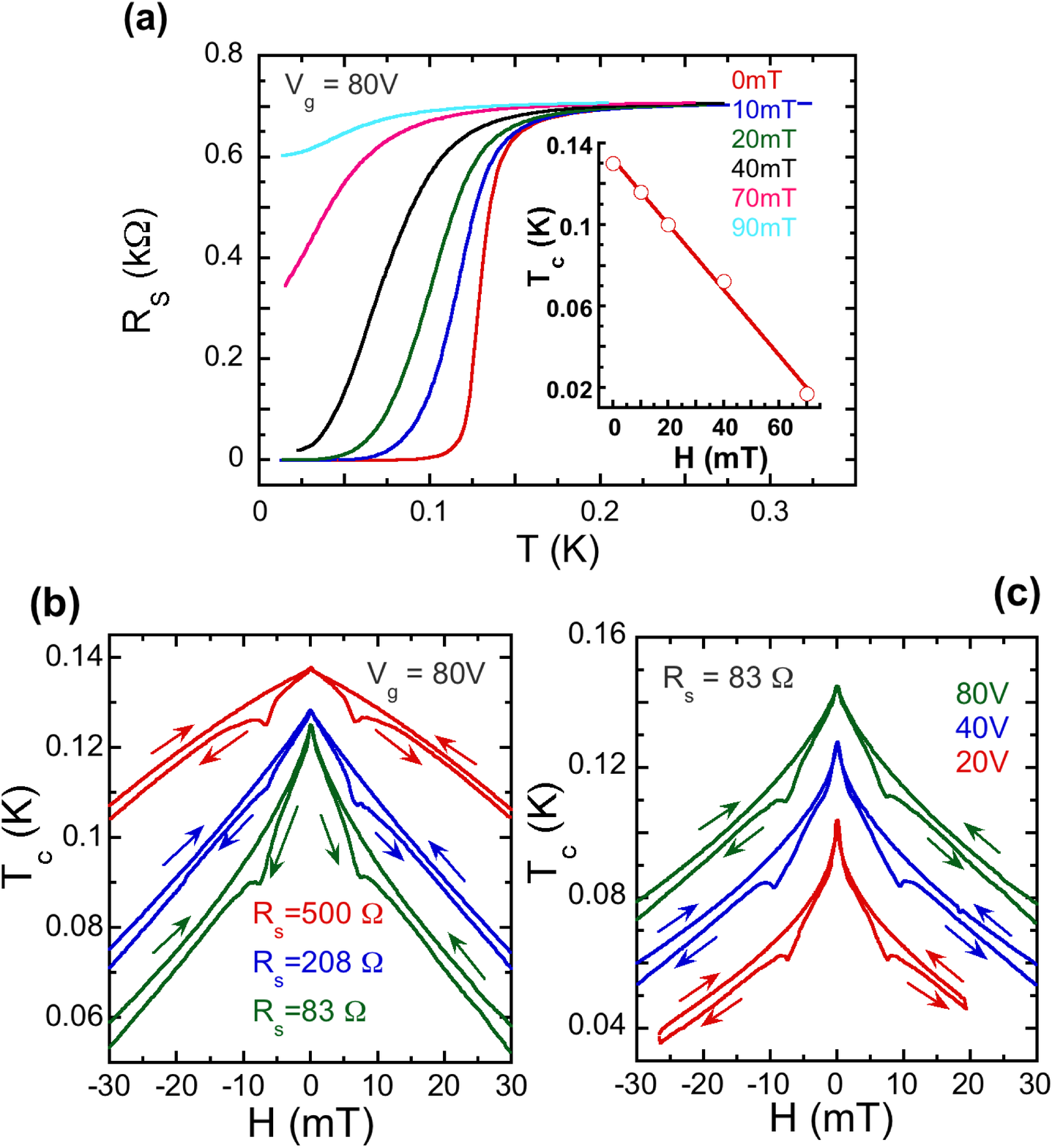}
\vspace{-0.5cm}
\caption{ (a)  Superconducting transition for $V_g=80$ V, at a few different magnetic fields.  
Inset: $T_c$ vs. $H$ obtained from these data.  $T_c$ is defined by the midpoint of the resistive transition.  The line represents a 
linear fit to the data points, with a slope of 1.6 mK/mT, giving a zero temperature coherence length of $\xi_0=64$ nm.  (b)  Phase diagram, 
$T_c$ vs. $H$, for a gate voltage of $V_g=80$ V, where the superconducting properties are maximal.  The three curves represent different 
resistance bias points along the superconducting transition, with the normal state resistance per square being
 704 $\Omega$.  
The arrows mark the direction of the magnetic field ramp.  (c)  Phase diagrams at a bias point of $R_S = 83$  $\Omega$, 
at the foot 
of the resistive transition, at three different gate voltages.   Data for $V_g=40$ V are shifted by 10 mK and data for $V_g=80$ V by 20 mK 
for clarity.  }
\vspace{-0.5cm}
\label{fig2}
\end{figure}   
Figure 2(a) shows the superconducting transition at $V_g=80$ V at a few different values of magnetic field $H$ applied 
perpendicular to the film plane.  Defining $T_c$ as the temperature corresponding to half the normal state resistance, a plot 
of $T_c(H)$ is shown as an inset to the figure.  For two dimensional superconductors in this field orientation, $T_c(H)$ should be 
linear, its slope proportional to the Ginzburg-Landau coherence length $\xi_{0}$ \cite{tinkham}.  A linear fit gives $\xi_0=$ 
64 nm, close to that reported previously \cite{reyren,shalom}.

However, this method of determining $T_c(H)$ misses some of the more novel and interesting behavior of 
this system.  To show this behavior, we map $T_c(H)$ continuously by controlling the sample temperature so that the 
sample resistance $R_S$ remains fixed as we ramp the magnetic field.  Figure 2(b) shows $T_c(H)$ for three different 
resistance bias points at $V_g$ = 80 V.  The most striking aspect of the data is that the behavior is hysteretic as a function of $H$.
Consider the curve corresponding to the bias point $R_S$ = 208 $\Omega$, slightly lower than the midpoint of the transition.  
Increasing the magnetic field from negative values towards $H=0$, $T_c$ shows a smooth increase that is almost linear.  The slope 
of this linear curve gives $\xi_0$ = 71 nm, close to the value obtained from Fig. 2(a).  On ramping $H$ beyond zero, however, 
the behavior becomes non-monotonic: in particular, there is a local minimum at $H\simeq$ 7 mT.  Increasing $H$ further, $T_c(H)$ 
becomes monotonic again.  Reversing the ramp direction results in a mirror image of the first trace, giving rise to a characteristic 
``butterfly'' curve.  Similar behavior is seen at other gate voltages for which the sample goes superconducting (Fig. 2(c)).    

Hysteretic behavior and butterfly curves in the magnetic field dependence of electrical characteristics are signatures of underlying 
ferromagnetic order in a sample.  Such behavior has already been reported in LAO/STO interface samples:  Brinkman \textit{et al.} \cite{brinkman} demonstrated that hysteretic behavior is observed at low 
temperatures in 26 uc thick LAO films grown under high $P_{O_2}$;
more recently, Ariando \textit{et al.} \cite{ariando} were able to observe hysteretic magnetization curves coexisting with paramagnetic and 
diamagnetic behavior in samples also grown in high $P_{O_2}$ that persisted to room temperature.   For low $P_{O_2}$ during growth, 
Huijben \textit{et al.} \cite{huijben} note that a low sheet resistance and metallic behavior is observed; for intermediate 
$P_{O_2}$, the samples go superconducting; while for high $P_{O_2}$ magnetic behavior is seen.  These experiments suggest to us that the magnetism is associated with the polar 
catastrophe mechanism, while the superconductivity is associated with the presence of oxygen vacancies.    However, in all previous 
experiments, the magnetic and superconducting regimes were 
quite distinct.  In contrast, our samples, which are grown under high $P_{O_2}$ and have $R_S$ consistent with the 
superconducting samples measured by others \cite{huijben}, show a coexistence of superconductivity and ferromagnetism.  The electronic properties of LAO/STO interfaces are extremely sensitive to growth conditions: samples grown by different groups with 
nominally identical growth conditions show some variations in electronic properties.  Our samples appear to be in a growth regime where both phenomena coexist.                

Caviglia \textit{et al.} \cite{caviglia} have reported weak (anti-)localization magnetoresistance (MR) in LAO/STO samples.  
At first sight, the data from our samples appear to be very similar to their data.  Figure 3(a) shows the MR of the sample 
at a few different 
temperatures at $V_g=-100$ V, not in the superconducting regime.  The magnitude of the resistance change and the sharpness of the 
resistance dip near zero field increase with decreasing temperature, consistent with a phase coherence length that increases with 
decreasing temperature.  
However, a closer look at the low field MR, shown in Fig. 3(b), reveals some significant differences from \cite{caviglia}.  In addition to the nonmonotonic MR over a large field scale (Fig. 3(a)), the samples show an additional 
resistance dip near zero field whose magnitude increases and width decreases with decreasing temperature.  This behavior, which is 
characteristic of weak localization (WL) \cite{bergmann}, suggests that there are two independent carrier gases that contribute to the conductance of 
the device in parallel, each with its own WL contribution.   The MR also shows a hysteretic ``butterfly'' pattern similar to that seen 
in $T_c(H)$ (Fig. 2(b,c)), indicating that local magnetic fields arising from magnetic order also modulate quantum 
interference in the carrier gases.  

\begin{figure}[h]
\includegraphics[width=9.0cm]{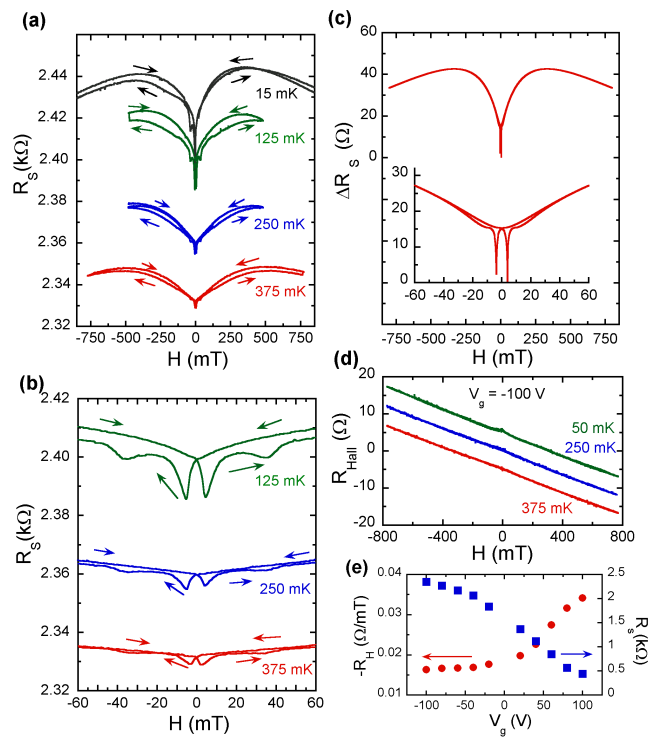}
\vspace{-0.5cm}
\caption{(a)  Magnetoresistance (MR) at $V_g=-100$ V at four different temperatures.  (b) 
Expansion of the low field data of (a) for the three highest temperatures.  (c)  Simulation of the longitudinal MR assuming weak 
localization (WL) contributions from two parallel conducting channels.  
 The inset shows an expanded view of the zero field region, showing the hysteresis and the non-monotonic dependence seen 
in the experimental data.  (d)  Hall resistance at $V_g=-100$V at three different temperatures.  Data for 50 mK and 375 mK have been 
offset by + 5 and -5 $\Omega$ respectively.  (e)  Hall coefficient and zero field longitudinal resistance at different gate voltages at 
270 mK.  The longitudinal resistance decreases with increasing gate voltage, but the Hall coefficient, nominally inversely proportional 
to the electron density, increases.   }
\vspace{-1.0cm}
\label{fig3}
\end{figure}  
To demonstrate the feasibility of this picture, we have simulated the WL contribution of two parallel two-dimensional 
gases in the presence of an external magnetic field and a magnetic field arising from a hysteretic intrinsic magnetization.  To calculate the WL contribution in the presence of spin-orbit scattering, we use the formalism of Santhanam \textit{et al.} \cite{santhanam}.  The resulting 
curve is shown in Fig. 3(c).  The simulation 
qualitatively reproduces the experimental features:  a non-monotonic MR over a large field scale, and a smaller non-monotonic MR over 
a smaller field scale, with hysteresis due to intrinsic magnetic order.  In the simulation, the larger contribution (88\%) is due to a carrier gas that has 
short phase coherence ($L_{\phi}$) and spin-orbit scattering ($L_{so}$) lengths ($L_{\phi} = 0.17$ $\mu$m, $L_{so} = 0.03$ $\mu$m), while the remaining contribution is due to a carrier 
gas with longer $L_{\phi}$ and $L_{so}$ ($L_{\phi} = 4.5$ $\mu$m, $L_{so} = 0.45$ $\mu$m).  Experimentally, we observe an increase in the amplitude of the low-field MR as the sample 
approaches the superconducting transition (by changing $V_g$, for example), which is probably due to the contribution of Maki-Thompson 
superconducting fluctuations \cite{maki-thompson} whose MR field scale is determined by $L_{\phi}$ \cite{santhanam}, and which is expected to increase exponentially 
as $T\rightarrow T_c$.  

Further evidence for two parallel conduction channels can be seen in the Hall resistance data.  Figure 3(d) 
shows the Hall resistance $R_{Hall}$ of the sample at $V_g=-100$ V at three different temperatures.  The sign of the slope 
of $R_{Hall}(H)$, $R_H$, corresponds to negatively charged carriers with a density $n \simeq 4.4 \times 10^{13}/$cm$^2$, assuming a 
single carrier band.  If there were only one electron gas at the interface, increasing $V_g$ should increase $n$, decreasing both $R_S$ and  $R_H$.  As shown in Fig. 3(e), increasing $V_g$ 
does indeed decrease $R_S$, but \textit{increases} $R_H$.  This behavior is only possible if there are at least 
two types of carriers involved in electrical transport, with different dependences of densities and mobilities on gate voltage.  Evidence for multiple charge carriers has also been observed by other groups \cite{seo,fix}.

\begin{figure}[!]
\includegraphics[width=7.0cm]{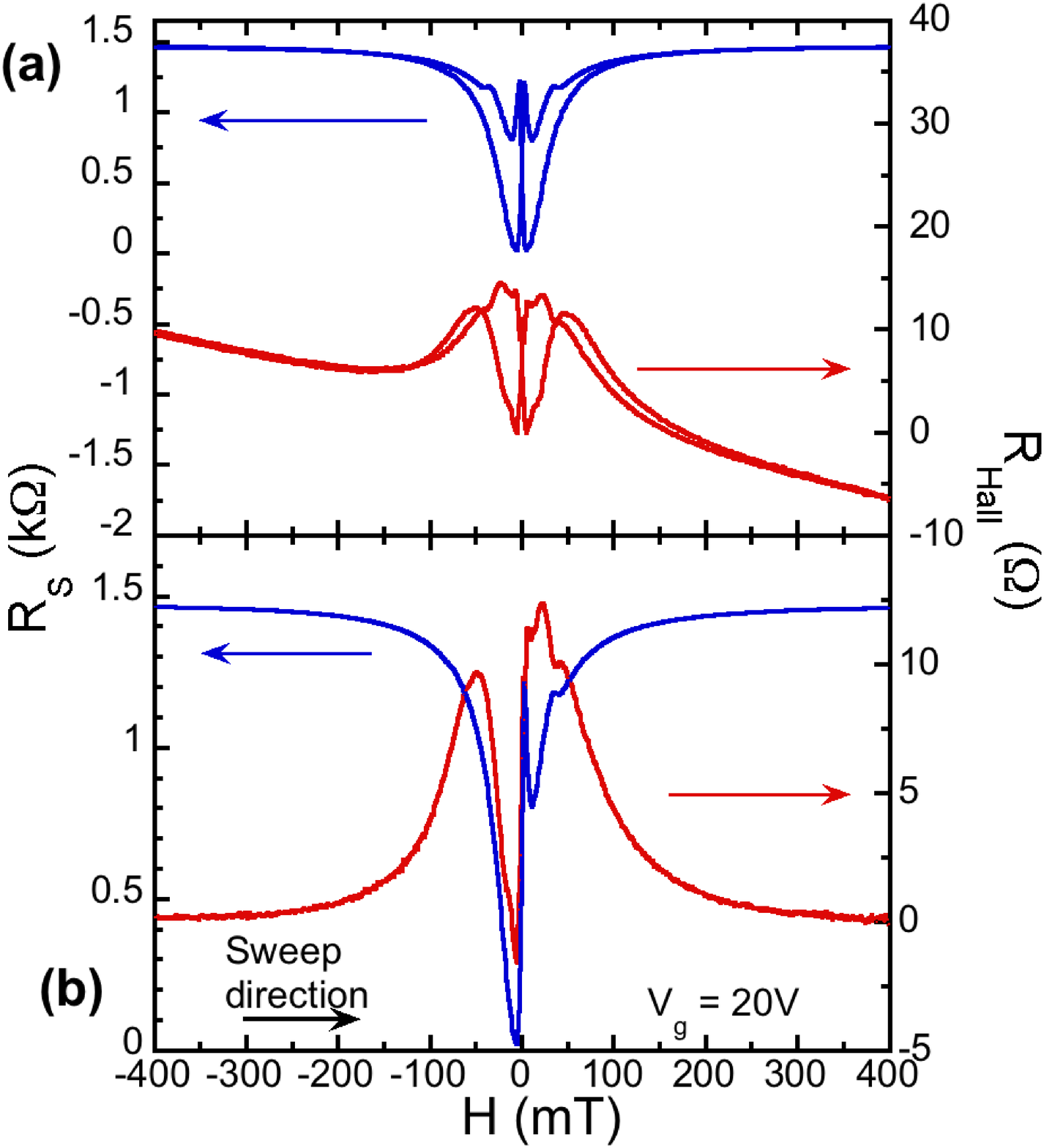}
\vspace{-0.3cm}
\caption{ (a)  Magnetoresistance (top curve) and Hall resistance (bottom curve) at $V_g=20$ V at 50 mK.  
(b) Same data as in (a), for one direction of the magnetic field sweep, except the background linear component is subtracted from the 
Hall resistance.     }
\vspace{-0.675cm}
\label{fig4}
\end{figure}
For conventional itinerant ferromagnets, it is well known that a finite magnetization 
should manifest itself as a contribution to $R_{Hall}$ through the anomalous Hall effect \cite{nagaosa}.  Figure 4(a) shows 
the longitudinal and Hall MR at $V_g=20$ V and 50 mK, where $R_S$=0 when $H$=0 (Fig. 1(a)).  Figure 4(b) shows the 
same data with the field sweep in one direction.  Here a linear contribution determined by fitting $R_{Hall}(H)$ at high magnetic fields 
has been subtracted.  It can be seen that some of 
the structure in $R_S(H)$ also has corresponding signatures in $R_{Hall}(H)$ (this structure is not due to misalignment of the Hall contacts, which is very small).  The structure in $R_S(H)$ and $R_{Hall}(H)$ for $H>0$ (for this field 
sweep direction) is also seen at other gate voltages, further from the superconducting transition.  We 
believe that this structure in the Hall resistance is due to an anomalous Hall effect arising from the interaction of the charge carriers 
with the magnetic moments at the interface, but the structure seen in Fig. 4(b) indicates that this interaction is more complicated than that in simple itinerant ferromagnets, and warrants further investigation.  
The features in $R_{Hall}(H)$ change with $V_g$.  The sharp dip in both the longitudinal and the Hall resistance for $H<0$ appears only 
close to the superconducting transition, and may be associated with vortex flow in the superconductor \cite{hagen}.

In summary, the interface between LAO and STO shows a rich variety of behavior,  including interacting ferromagnetism and superconductivity.  Our measurements indicate that there are two different types of charge carriers in the system.  The ability to tune 
the properties of the system by means of a gate voltage makes this a fascinating 
system for studying competing cooperative phenomena in two dimensions. 

We thank J.B. Ketterson and A.J. Freeman for useful discussions.  This work was funded through a grant from the US Department of Energy through grant number DE-FG02-06ER46346.

\newpage

\newpage

\end{document}